\documentclass[letterpaper]{article}

\usepackage{arxiv}
\usepackage{algorithm}
\usepackage{algpseudocode}
\usepackage{amsfonts}
\usepackage{amsmath}
\usepackage{amssymb}
\usepackage{amsthm}
\usepackage{bbm}
\usepackage{booktabs}
\usepackage{dsfont}
\usepackage{mathrsfs}
\usepackage{multirow}
\usepackage{pgfplots}
\usepgfplotslibrary{fillbetween}
\usepackage{tikz}
\usepackage{siunitx}
\usepackage{stmaryrd}
\usepackage{subcaption}
\usepackage{xcolor}
\usepackage{hyperref}
\usepackage{varwidth}
\usepackage[normalem]{ulem}
\usepackage{appendix}
\usepackage{bm}
\usepackage{enumitem}

\usepgfplotslibrary{external} % compile figures outside of the main document and include them as pdf
\tikzsetexternalprefix{ext/}

\makeatletter
\g@addto@macro\@floatboxreset\centering
\makeatother

\usetikzlibrary{calc,fadings,shapes.misc,3d,arrows,
                decorations,decorations.pathmorphing,
                decorations.text,shapes.geometric,
                decorations.markings,patterns,spy,positioning,
                arrows.meta,quotes}

\usepackage{tikz-3dplot}
\definecolor{myblue1}{RGB}{0,177,234}
\definecolor{myblue2}{RGB}{76,200,239}
\definecolor{myblue3}{RGB}{127,215,244}
\definecolor{myblue4}{RGB}{178,231,248}
\definecolor{mybluegray1}{RGB}{0,127,167}
\definecolor{mybluegray2}{RGB}{76,165,193}
\definecolor{mybluegray3}{RGB}{127,191,211}
\definecolor{mybluegray4}{RGB}{178,216,228}
\definecolor{mygray1}{RGB}{76,84,93}
\definecolor{mygray2}{RGB}{129,135,141}
\definecolor{mygray3}{RGB}{165,169,174}
\definecolor{mygray4}{RGB}{201,203,206}
\definecolor{mygray5}{RGB}{231,233,236}
\definecolor{myorange1}{RGB}{255,126,46}
\definecolor{myorange2}{RGB}{255,164,108}
\definecolor{myorange3}{RGB}{255,190,150}
\definecolor{myorange4}{RGB}{255,216,192}

\newcommand{\eps}{\varepsilon}

\newcommand{\V}[1]{\textbf{#1}}
\newcommand{\GV}[1]{\boldsymbol{#1}}

\newcommand{\ie}{\textit{i.e.}\;}
\newcommand{\etal}{\textit{et al}\;}

\newcommand{\tb}[1]{\textcolor{black}{#1}}

\setenumerate{label=\textcolor{mybluegray1}{$\bm{\diamond}$},itemsep=1pt,topsep=5pt} 
\title{Direct shape optimization through deep reinforcement learning}

\author{
	Jonathan Viquerat\thanks{Corresponding author} \\
	MINES Paristech , CEMEF\\
	PSL - Research University\\
	\texttt{jonathan.viquerat@mines-paristech.fr} \\
\And
	Jean Rabault\\
	Department of Mathematics\\ 
	University of Oslo
\And
	Alexander Kuhnle\\
	University of Cambridge
\And
	Hassan Ghraieb \\
	MINES Paristech , CEMEF\\
	PSL - Research University
\And
	Aur\'elien Larcher \\
	MINES Paristech , CEMEF\\
	PSL - Research University
\And
	Elie Hachem \\
	MINES Paristech , CEMEF\\
	PSL - Research University
}
\begin{document}
\maketitle

\begin{abstract}
Deep Reinforcement Learning (DRL) has recently spread into a range of domains
within physics and engineering, with multiple remarkable
achievements. Still, much remains to be explored before the capabilities of
these methods are well understood. In this paper, we present the first
application of DRL to direct shape optimization. We show that, given adequate reward, an
artificial neural network trained through DRL is able to generate optimal
shapes on its own, without any
prior knowledge and in a constrained time. While we choose here to apply this
methodology to aerodynamics, the optimization process
itself is agnostic to details
of the use case, and thus our work paves the
way to new generic shape optimization strategies both in fluid mechanics,
and more generally in any domain where a relevant reward function
can be defined.
\end{abstract}


%%%%%%%%%%%%%%%%%%%%%%%%%%%%%%%%%%%%%%%%%%%%%%%%%%%%%%%%%%
%%%%%%%%%%%%%%%%%%%%%%%%%%%%%%%%%%%%%%%%%%%%%%%%%%%%%%%%%%
%%%%%%%%%%%%%%%%%%%%%%%%%%%%%%%%%%%%%%%%%%%%%%%%%%%%%%%%%%
\section{Introduction}

\noindent Shape optimization is a long-standing research topic with countless industrial
applications, ranging from structural mechanics to electromagnetism and biomechanics \cite{Allaire1997} \cite{Semmler2015}.
In fluid dynamics, the interest in shape optimization has been driven by many real-world
problems. For example, within aerodynamics, the reduction of drag and therefore of fuel consumption by trucks and
cars \cite{hucho1993aerodynamics}, or
the reduction of aircraft fuel consumption and running costs \cite{Jameson2003}, are cases on which a large body of
literature is available. However, shape optimization also plays a key role in many other aspects of
the performance of, for example, planes, and modern optimization techniques are also applied to a variety of
problems such as the optimization of electromagnetically
stealth aircrafts \cite{Jang2016}, or acoustic noise reduction
\cite{Marsden2001}. This illustrates the importance of shape optimization methods in many
applications, across topics of both academic and industrial interest.

In the following, we will focus on one shape optimization problem, namely airfoil shape optimization.
This problem is key to many industrial processes, and presents the ingredients of non-linearity
and high dimensionality which make shape optimization at large a challenging problem.
As a consequence of its industrial and academic relevance, airfoil shape optimization through numerical
techniques has been discussed since at least
back to 1964 \cite{Carlson1964}, and remains a problem of active research
\cite{Reuther2013, Zang2010}.

Following developments in optimization techniques, two main classes of approaches have emerged to tackle
shape optimization problems, namely gradient-based and
gradient-free methods. Gradient-based methods rely on the evaluation of
$\nabla_x J$, the gradient of the objective function $J$ with respect to
the design parameters $x$. These methods have been widely used for their
low computational cost in large optimization spaces
\cite{Kenway2015}, where the gradient computation by adjoint methods has
proven to be very efficient \cite{Pironneau1974}. The major drawbacks of
gradient-based methods are (i) that they can easily get trapped in local
optima and are therefore highly sensitive to the provided starting point,
especially when strongly nonlinear systems are studied, and (ii) that their efficiency is severely
challenged in situations where the objective function exhibits discontinuities
or is strongly non-linear \cite{Skinner2018}. Gradient-free
methods are clearly superior in this context, however, their
implementation and application can be more complex. Among gradient-free methods,
genetic algorithms \cite{Yamazaki2008} are known to be good at finding global optima,
and are also less sensitive to computational noise than gradient-based methods.
However, their computational cost is usually higher than gradient-based methods, thus limiting the number
of design parameters the method can tackle \cite{Skinner2018}. Particle swarm
optimization is another well-known method praised for its easy implementation
and its low memory cost \cite{Wang2011}. Its major drawback is the difficulty
to impose constraints on the design parameters \cite{Hassan2012}. A last major
class of gradient-free algorithms are the metropolis algorithms, such as
simulated annealing \cite{Tiow2002}. This method, based on the physical process
of molten metal cooling, is well-known for its ability of escaping local
minima, although the results obtained can be highly dependent on the chosen
meta-parameters of the algorithm. Finally, it should be noted that for both
gradient-based and gradient-free methods, a surrogate model can be used for the computational part,
instead of systematically relying on a CFD solver \cite{Koziel2016}. Many
methods to construct such surrogate models exist, such as radial basis
functions, kriging, or supervised artificial neural networks \cite{Queipo2005}.
In all these methods, the geometric parameterization plays a determinant
role, both for the attainable geometries and for the tractability of the
optimization process \cite{Chernukhin2013}. In particular, parameterizations
based on B\'ezier curves \cite{Peigin2008}, B-splines \cite{Koziel2016} and
NURBS \cite{DellaVecchia2014} have been widely studied within conventional
optimization frameworks.

\begin{figure}[h!]
\centering
%%%%%%%
\includegraphics{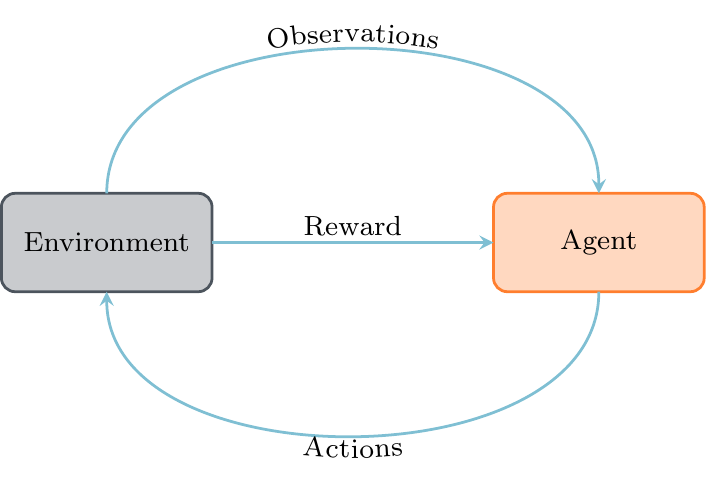}
%%%%%%%
\caption{\textbf{General reinforcement learning framework}. }
\label{fig:DRL}
\end{figure}

As of today, the use of supervised neural networks in conjunction with
gradient-based and gradient-free methods for shape optimization is supported by
an abundant literature. In supervised learning, a labeled dataset (\ie pairs
of inputs and expected outputs) is used to train a neural network until it
approximates the mapping function between the input and output spaces accurately \cite{Goodfellow2017}.
\tb{Several such approaches for computation fluid dynamic problems can be found in the review \cite{Brunton2020}.}
However, \tb{at the time of writing}, almost no work exploiting neural networks in combination with
reinforcement learning (RL) \tb{could be found} for similar goals. In RL,
an agent interacts with an environment in a closed
loop, as depicted in figure \ref{fig:DRL}. At each time $t$ in the interaction, the agent
(here, the neural network) is provided with a partial observation of the environment
state $s_t$, and in response outputs an action $a_t$ to be
executed, which influences the subsequent evolution of the environment.
Moreover, the agent periodically receives a reward signal $r_t$ reflecting the
quality of the actions recently taken, and the goal of RL is to find an optimal
decision policy $a_t = \pi(s_t)$ maximizing its cumulated reward. \tb{Video games provide a good conceptual comparison: the environment (the game) provides observations (the screen, the sounds) to the player (the agent), which in response provides actions (through the controller) to the game, thus modifying the current state of the game. The reward usually consists of the score attained by the player. The latter tries to increase its overall score by learning the optimal actions to take given the observations provided by the game.} As RL is mostly
agnostic to details of the environment, any time-dependent process satisfying
the required state/action/reward interface is eligible, ranging from games like Atari 
\cite{Mnih2013}, Go \cite{silver2017mastering} or
poker \cite{Browneaay2400}, to
physics simulations \cite{Heess2017, Rabault2018} or
resource management problem \cite{Mao2016}.

%%%%%%%%%%%%%%%%%%%%
%%%%%%%%%%%%%%%%%%%%

%\input{fig/DRL}

A recent review presented the available literature on DRL for fluid 
dynamics applications \cite{Garnier2019}. This work highlighted the 
potential of DRL in the context of fluid mechanics.
To the best knowledge of the authors, \tb{at the time of writing}, only two contributions proposed to
exploit RL for problems related to shape optimization. These papers are
concerned with the optimization of morphing airfoils with two
\cite{Lampton2010} and four \cite{Lampton2008} parameters, respectively, using
Q-learning methods (see section \ref{section:methodology}). In these
contributions, the agent discovers the optimal morphing dynamics of airfoils in
transition between different flight regimes.
In addition, although not focusing on shape optimization, it is worth noting the work
presented in \cite{Andrychowicz2016}, where neural networks directly learn how
to perform gradient descent on specific classes of problems. The method is
reported to have a very good generalization capability, which confirms it as a good
candidate for shape optimization. Therefore, it appears that while DRL is a promising
method for performing shape optimization, there has not been a clear illustration of
direct shape optimization with DRL yet.

Therefore, motivated by our observation of the potential of DRL for performing
shape optimization, we present in this article the first application of deep reinforcement
learning (DRL) to direct shape optimization. We use Proximal Policy
Optimization (PPO \cite{Schulman2017}) in combination with an artificial neural network
to generate 2D shapes described by B\'ezier curves. The flow around the shape
is assessed via 2D numerical simulation at moderate Reynolds numbers using
FeniCs \cite{Fenics2015}. We commit to releasing all our code as open source, in order
to facilitate the application of DRL to shape optimization. In the following,
we first describe the methodology,
including the shape generation method and the numerical setup used to
compute the flow around the generated shapes. Subsequently, we introduce our
reinforcement learning approach and its application to shape optimization.
Finally, several results of the optimization process are presented.

%%%%%%%%%%%%%%%%%%%%%%%%%%%%%%%%%%%%%%%%%%%%%%%%%%%%%%%%%%
%%%%%%%%%%%%%%%%%%%%%%%%%%%%%%%%%%%%%%%%%%%%%%%%%%%%%%%%%%
%%%%%%%%%%%%%%%%%%%%%%%%%%%%%%%%%%%%%%%%%%%%%%%%%%%%%%%%%%
\section{Methodology}
\label{section:methodology}

%%%%%%%%%%%%%%%%%%%%%%%%%%%%%%%%%%%%%%%%%%%%%%%%%%%%%%%%%%
%%%%%%%%%%%%%%%%%%%%%%%%%%%%%%%%%%%%%%%%%%%%%%%%%%%%%%%%%%
\subsection{CFD environment}
\label{section:cfd}

\begin{figure}
\centering
%%%%%%%
\begin{subfigure}[t]{.45\textwidth}
	\centering
	\fbox{\includegraphics[width=\linewidth]{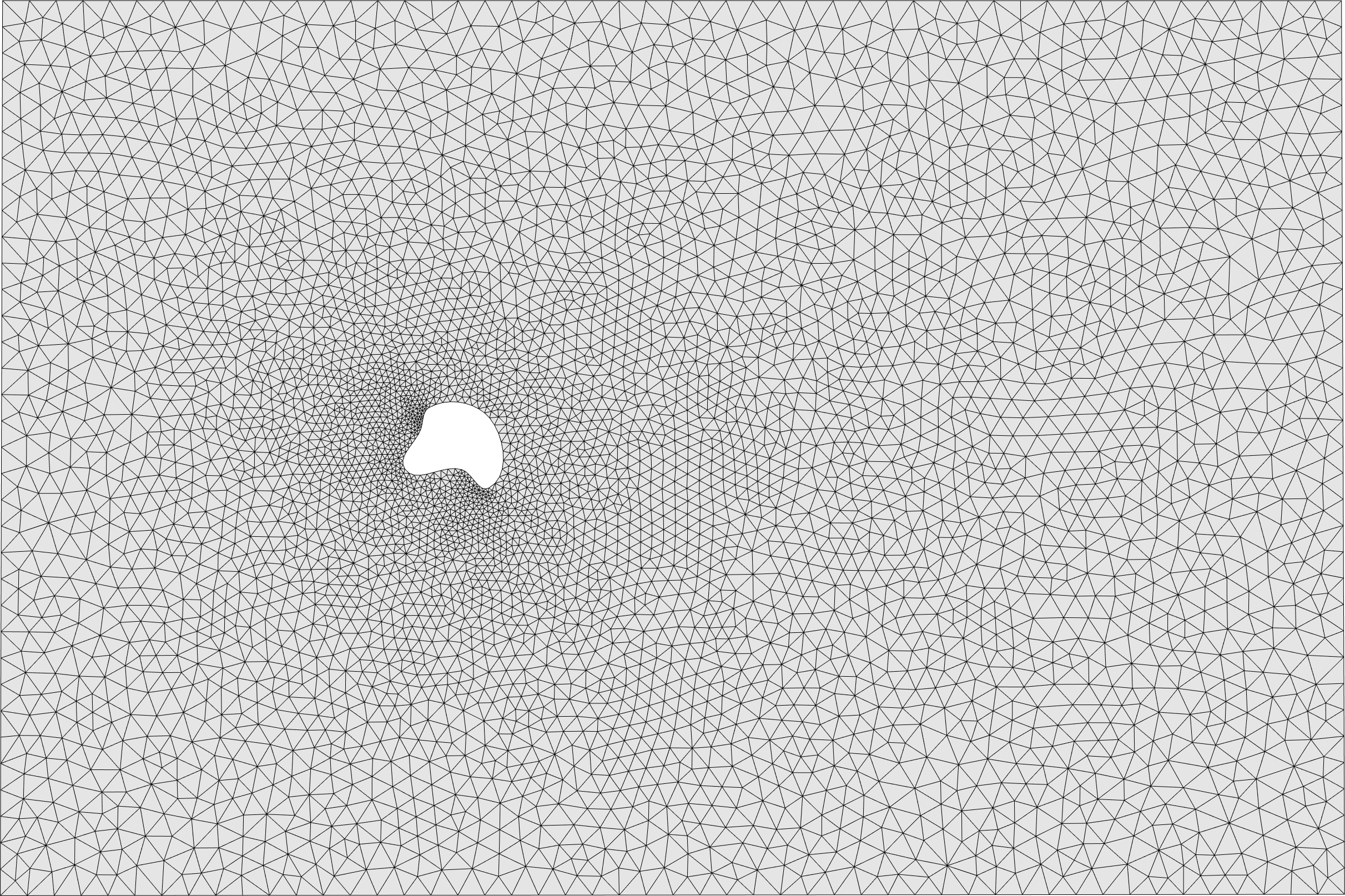}} 
	\caption{Mesh of the computational domain}
	\label{fig:domain_mesh}
\end{subfigure} \qquad
\begin{subfigure}[t]{.45\textwidth}
	\centering
	\fbox{\includegraphics[width=\linewidth]{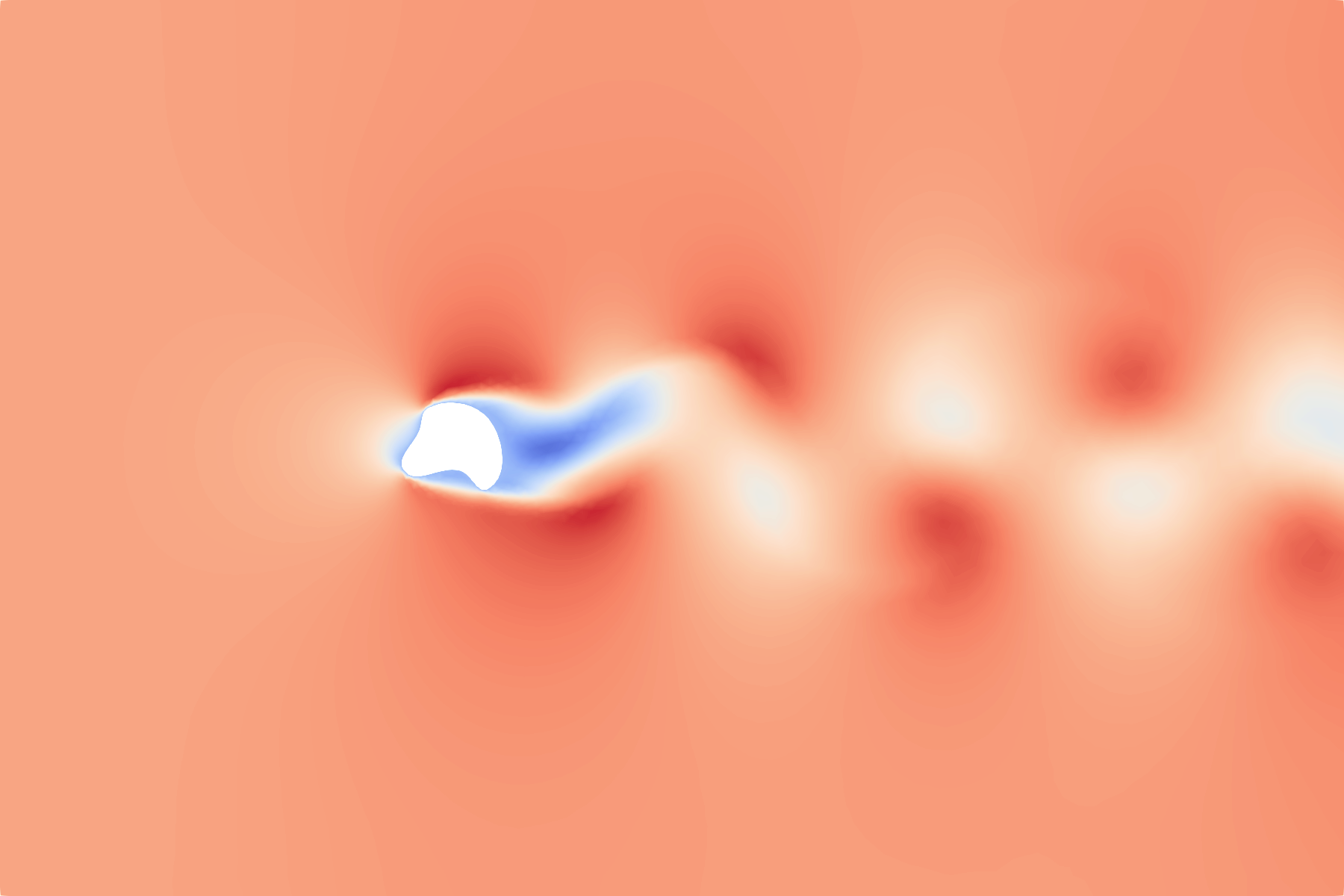}}
	\caption{Computed $v_x$ velocity field at $Re=200$}
	\label{fig:domain_sol}
\end{subfigure}
%%%%%%%
\caption{\textbf{Mesh and computed $v_x$ velocity field at $Re=200$} behind a random shape. The recirculation area behind the obstacle is clearly visible, followed by a well-established Von Karman vortex alley. The velocity field is scaled in the $\left[-1, 1\right]$ range.}
\label{fig:domain}
\end{figure}

The CFD simulation, which constitutes the environment interacting with the
DRL agent, consists of a computational fluid dynamics (CFD) simulation
based on FeniCs \cite{Fenics2015} that numerically solves the Navier-Stokes (NS)
equations. Each shape, of typical dimension 1, is immersed in a rectangular
computational domain of length $l=45$ and width $w=30$ (see figure
\ref{fig:domain_mesh}). A constant velocity $\V{v}=v_{\text{in}} \V{e}_x$ is
imposed on the inflow profile, while free-slip boundary conditions are imposed
on top and bottom of the domain. Finally, a no-slip boundary condition is
imposed on the obstacle, and \tb{a traction-free} condition is set on the outflow
profile. To perform the required numerical computations, mesh generation of the
domain and geometries is performed using Gmsh \cite{Geuzaine2017}. The
reference flow corresponds to that of a cylinder of radius $r_{\text{cyl}} = 1$
immersed in the same domain. The reference Reynolds number is then defined as:

\begin{equation}
	Re_\text{ref} = \frac{2 \rho v_{\text{in}} r_{\text{cyl}}}{\mu},
\end{equation}

\noindent where $\rho$ is the volumic mass of fluid, and $\mu$ its viscosity.
In the remaining of this paper, $\rho$ is kept constant and equal to
1 kg/m${}^3$, as well as $v_{\text{in}}$, which is kept equal to 1 m/s. The
modification of flow conditions is performed through the choice of the
reference Reynolds number, which is modified by adapting the viscosity of the fluid.
For all computations, the numerical timestep is chosen as: 

\begin{equation}
	\Delta t = C \, \frac{h_{\text{min}}}{v_{\text{in}}},
\end{equation}

\noindent the constant $C$ being the CFL condition number (here, $C=0.5$).  The drag and lift
forces experienced by a given shape immersed in the flow are computed as
follows:

\begin{equation}
	f_{d} = \int_{S} \left( \GV{\sigma} \cdot \V{n} \right) \cdot \V{e}_{x} \; \text{ and } \; f_{l} = \int_{S} \left( \GV{\sigma} \cdot \V{n} \right) \cdot \V{e}_{y},
\end{equation}

\noindent after what the drag and lift coefficients $C_d$ and $C_l$ are evaluated as:

\begin{equation}
	C_d = \frac{f_d}{\frac{1}{2} \rho v_{\text{in}}^2 s} \; \text{ and } \; C_l = \frac{f_l}{\frac{1}{2} \rho v_{\text{in}}^2 s}.
\end{equation}

In the following, a positive value of $C_d$ (resp $C_l$) indicates
that the force $f_d$ (resp $f_l$) is oriented toward $\V{e}_x$ (resp
$\V{e}_y$). The maximal physical time $t_\text{max}$ used in the numerical computations is
set in order to obtain stabilized average values of the monitored quantities of
interest (see next section). In practice, the following rule of thumb is used:

\begin{equation}
	t_\text{max} = \frac{2}{v_{\text{in}}} \left( x_{\text{max}} - x_{\text{min}} \right).
\end{equation}

The numerical formulation used for the resolution of the discretized
Navier-Stokes equations is a finite-element incompressible solver based on
the projection method \cite{Guermond2006}, coupled to a BDF2 marching-in-time
scheme. This allows us to consider flows at low Reynolds numbers, typically $Re = 200$.
Using this typical value of the Reynolds number allows to address a shape
optimization task presenting the ingredients of non linearity and high dimensionality
that are challenging in this class of problems, while keeping the computational budget
limited, therefore, allowing relatively fast training without large computational
resources. This is a similar approach to what was used in \cite{Rabault2018}, and is well
suited for a proof-of-concept of the methodology as well as future benchmarking
of fine-tuned algorithms.

%%%%%%%%%%%%%%%%%%%%
%%%%%%%%%%%%%%%%%%%%

%\input{fig/domain}

%%%%%%%%%%%%%%%%%%%%%%%%%%%%%%%%%%%%%%%%%%%%%%%%%%%%%%%%%%
%%%%%%%%%%%%%%%%%%%%%%%%%%%%%%%%%%%%%%%%%%%%%%%%%%%%%%%%%%
\subsection{Deep reinforcement learning}
\label{section:DRL}

Our DRL agent is based on proximal policy optimization (PPO) introduced in
\cite{Schulman2017}. This algorithm belongs to the class of policy gradient
methods, which differs from action-value methods such as Q-learning. In
action-value methods, the neural network learns to predict the future reward of
taking an action when provided a given state, and selects the maximum action based on
these estimates. In contrast, policy gradient methods directly optimize the
expected overall reward of a decision policy mapping states to actions, without
resorting to a value function. Policy gradient methods are advantageous since
(i) they can handle continuous action spaces (different from Q-learning
methods), and (ii) they were successfully applied to optimal control in a
similar setup in the past \cite{Rabault2018}.

Policy gradient methods are trained based on episodes, which correspond to a
sequence of consecutive interactions of the agent with the environment. The
temporally discounted sum of rewards observed during an episode acts as
the quality assessment of the agent's current decision policy. This in turn is used as
the learning signal to update the network parameters via gradient descent
\cite{Schulman2017}. By repeating this process, the agent iteratively learns to
choose more adequate actions in order to maximize its reward (details regarding
the PPO algorithm are given in appendix \ref{appendix:ppo}).

In this work, the network action output consists of $3n$ values in $\left[-1,
1\right]$, where $n$ is fixed for each experiment and corresponds to the number of points used to
specify the shape. These values are
then transformed adequately in order to generate a valid shape, by generating the
position and local curvature of a series of points connected through a Bezier curve. Given a $(p, q,
s)$ triplet provided by the network, a transformed triplet $(x, y, e)$ is
obtained that generates the position $x, y$ and local curvature $e$ of the
$i^{th}$ point:

\begin{equation}
	\left\{
	\begin{aligned}
		r &= r_{\text{max}} \max \left( \left| p \right|, r_{\text{min}} \right),\\
		\theta &= \frac{\pi}{n} \left( i + \frac{q}{2} \right),\\
		x &= r \cos \left( \theta \right),\\
		y &= r \sin \left( \theta \right),\\
		e &= \frac{1}{2} \left(1 + s \right).
	\end{aligned}
	\right.
\end{equation}

\begin{figure}[t]
\centering
%%%%%%%
\fbox{\includegraphics[width=.4\linewidth]{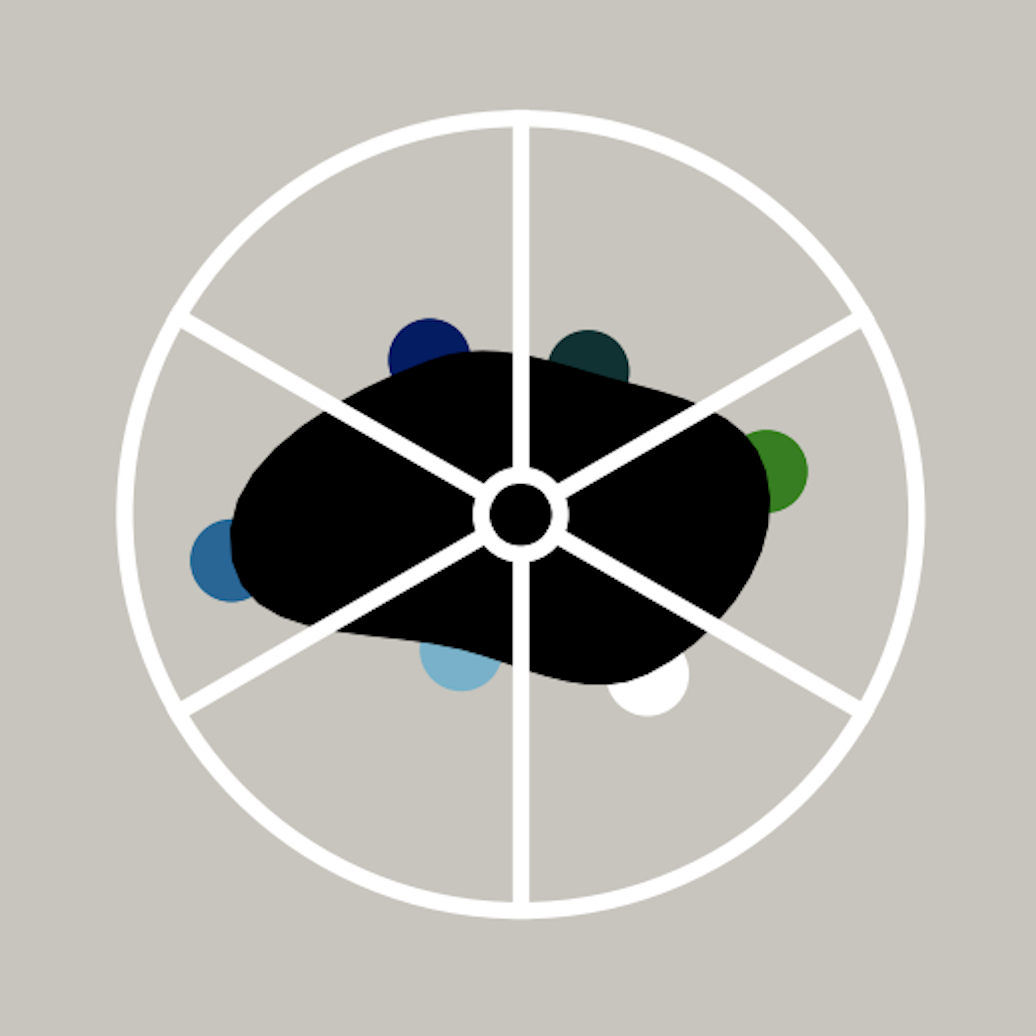}} 
%%%%%%%
\caption{\textbf{Example of generated shape with geometrical constraints.} The colored dots indicate the control points generated by the agent, that are then joined using B\'ezier curves. Each point the agent suggests is restricted by construction in radius (inner and outer circles of radii $r_{\text{min}}$ and $r_{\text{max}}$) and in azimuth (diverging white lines).}
\label{fig:shape}
\end{figure}

This mapping allows to restrict the reachable positions of each point
to a specific portion of a torus, bounded by user-defined interior
($r_{\text{min}}$) and exterior ($r_{\text{max}}$) radii (see figure
\ref{fig:shape}). In doing so, it encourages the generation of untangled shapes, thus
limiting resulting meshing issues. Once the final point positions are
computed, the environment connects them to generate a closed shape using
B\'ezier curves in a fully deterministic way (more details about the shape
generation process can be found in appendix \ref{appendix:shapes}). A CFD run is performed as described in
section \ref{section:cfd}, and once finished, the reward is
computed and passed on to the agent. The neural network representing the agent is a simple
fully-connected network with two hidden layers of size 512, similar to the
choice of \cite{Rabault2018}. Our training setup also benefits from
parallelized multi-environment training \cite{Rabault2019}, which provides an
almost-linear speedup with respect to the number of available cores.

%%%%%%%%%%%%%%%%%%%%
%%%%%%%%%%%%%%%%%%%%

%\input{fig/shape}

%%%%%%%%%%%%%%%%%%%%%%%%%%%%%%%%%%%%%%%%%%%%%%%%%%%%%%%%%%
%%%%%%%%%%%%%%%%%%%%%%%%%%%%%%%%%%%%%%%%%%%%%%%%%%%%%%%%%%
\subsection{Degenerate DRL}

\begin{figure}[b]
\centering
%%%%%%%
\includegraphics[width=\linewidth]{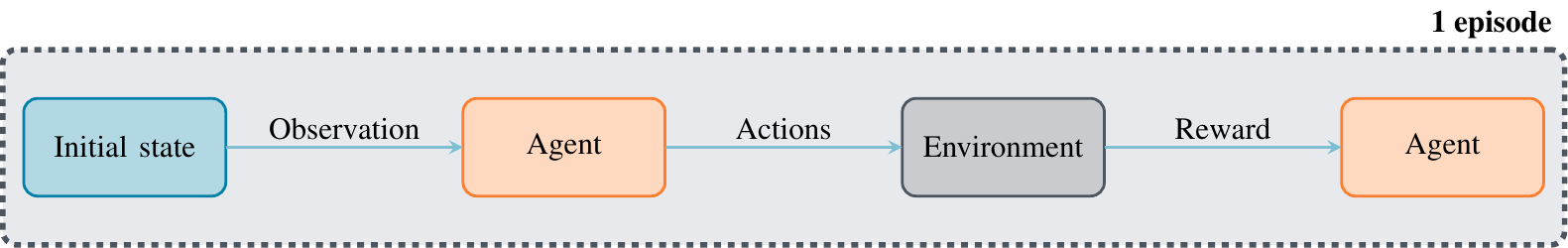} 
%%%%%%%
\caption{\textbf{Degenerate reinforcement learning framework used in this paper}. One episode consists of a single control from the agent: the same initial observation is provided to the agent at the start, which in return provides an action to the environment. The environment returns a reward value to the agent, and the episode is terminated.}
\label{fig:degenerate_DRL}
\end{figure}

For this paper, our setup follows a "degenerate" version of DRL where a
learning episode only consists of a single timestep, \ie a single attempt of
the network to generate an optimal shape (see figure \ref{fig:degenerate_DRL}).
Consequently, we only leverage the capability of DRL to learn from indirect
supervision through a generic reward signal (note that the correct optimal
response is not known, so supervised methods are not straightforwardly applicable).
As will be shown in section \ref{section:results}, this
choice allows to exploit DRL algorithms as direct non-linear optimizers. We are not
aware of other works applying DRL this way. %In future work, we plan to extend our setup
%to multi-timestep episodes, by interpreting a sequence of actions 
%as iterative refinements of an initial suboptimal shape.

%%%%%%%%%%%%%%%%%%%%
%%%%%%%%%%%%%%%%%%%%

%\input{fig/degenerate_DRL}

%%%%%%%%%%%%%%%%%%%%%%%%%%%%%%%%%%%%%%%%%%%%%%%%%%%%%%%%%%
%%%%%%%%%%%%%%%%%%%%%%%%%%%%%%%%%%%%%%%%%%%%%%%%%%%%%%%%%%
%%%%%%%%%%%%%%%%%%%%%%%%%%%%%%%%%%%%%%%%%%%%%%%%%%%%%%%%%%
\section{Results}
\label{section:results}

For the record, we are concerned with the generation of shapes maximizing the 
lift-to-drag ratio $\frac{C_l}{C_d}$. To this end, the proposed baseline reward is:

\begin{equation}
\label{eq:reward_wing}
	r_t = \left< \frac{C_l}{\left| C_d \right|} \right> - \left< \frac{C_l}{\left| C_d \right|} \right>_{\text{cyl}},
\end{equation} 

\noindent where the notation $<\cdot>$ indicates a temporal average over the
second half of the CFD computation. The ${}_{\text{cyl}}$ subscript corresponds
to the value computed in the reference case, \ie using the unit radius
cylinder. Here, the reference lift-to-drag value using the cylinder is
trivially equal to 0, the cylinder generating no lift in average. The fact that this
reward is sign-changing depending on the direction in which lift occurs implies
a good reward variability, helping the agent to learn. Finally, shapes where no reward 
could be computed (meshing failed or CFD computation failed) are penalized \textit{via} 
the reward function as follows:

\begin{equation}
\label{eq:reward_penalize_wing}
	r_t \leftarrow \max \left( r_t, r_{\text{fail}} \right).
\end{equation} 

This shaping also allows to clip the reward in the case of shape with
very bad aerodynamic properties. In practice, $r_{\text{fail}} = -5$. 
The deformations bounds are set as $r_{\text{min}} = 0.3$ 
and $r_{\text{max}} = 3$. The network parameters are updated every 50 shapes, 
with a learning rate equal to \num{1e-3}. The shape is described with 4 points, 
with the possibility of keeping
some of them fixed. 

%%%%%%%%%%%%%%%%%%%%%%%%%%%%%%%%%%%%%%%%%%%%%%%%%%%%%%%%%%
%%%%%%%%%%%%%%%%%%%%%%%%%%%%%%%%%%%%%%%%%%%%%%%%%%%%%%%%%%
\subsection{Baseline results}

\begin{figure}
\centering
%%%%%%%
\begin{subfigure}[t]{.25\textwidth}
	\centering
	\fbox{\includegraphics[width=\linewidth]{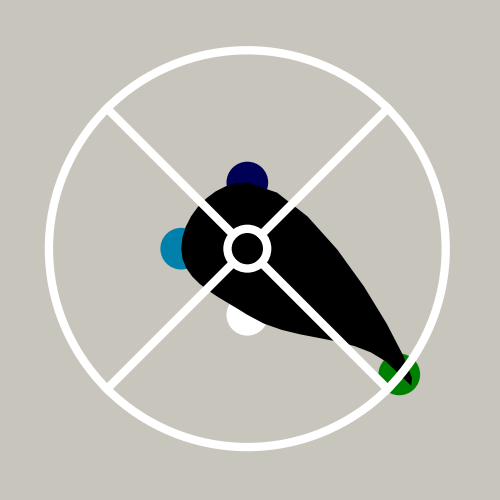}} 
	\caption{Best shape with 4 points, 1 free point (3 d.o.f.s)}
	\label{fig:4_pts_1_pt}
\end{subfigure} \qquad
\begin{subfigure}[t]{.25\textwidth}
	\centering
	\fbox{\includegraphics[width=\linewidth]{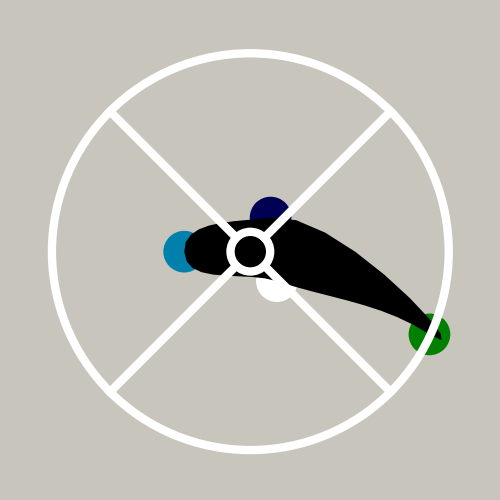}} 
	\caption{Best shape with 4 points, 3 free points (9 d.o.f.s)}
	\label{fig:4_pts_3_pts}
\end{subfigure} \qquad
\begin{subfigure}[t]{.25\textwidth}
	\centering
	\fbox{\includegraphics[width=\linewidth]{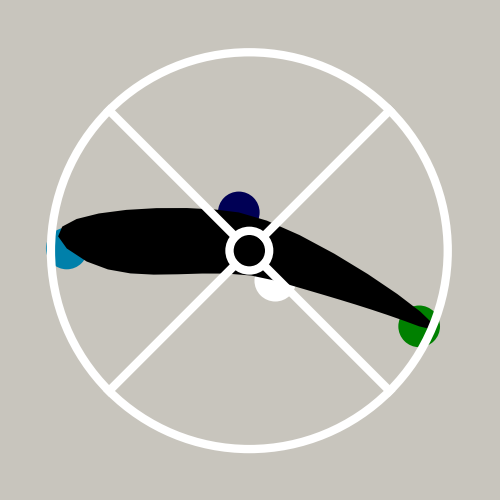}} 
	\caption{Best shape with 4 points, 4 free points (12 d.o.f.s)}
	\label{fig:4_pts_4_pts}
\end{subfigure}

\bigskip

\begin{subfigure}[t]{.875\textwidth}
	\centering
	\fbox{\includegraphics[width=\linewidth]{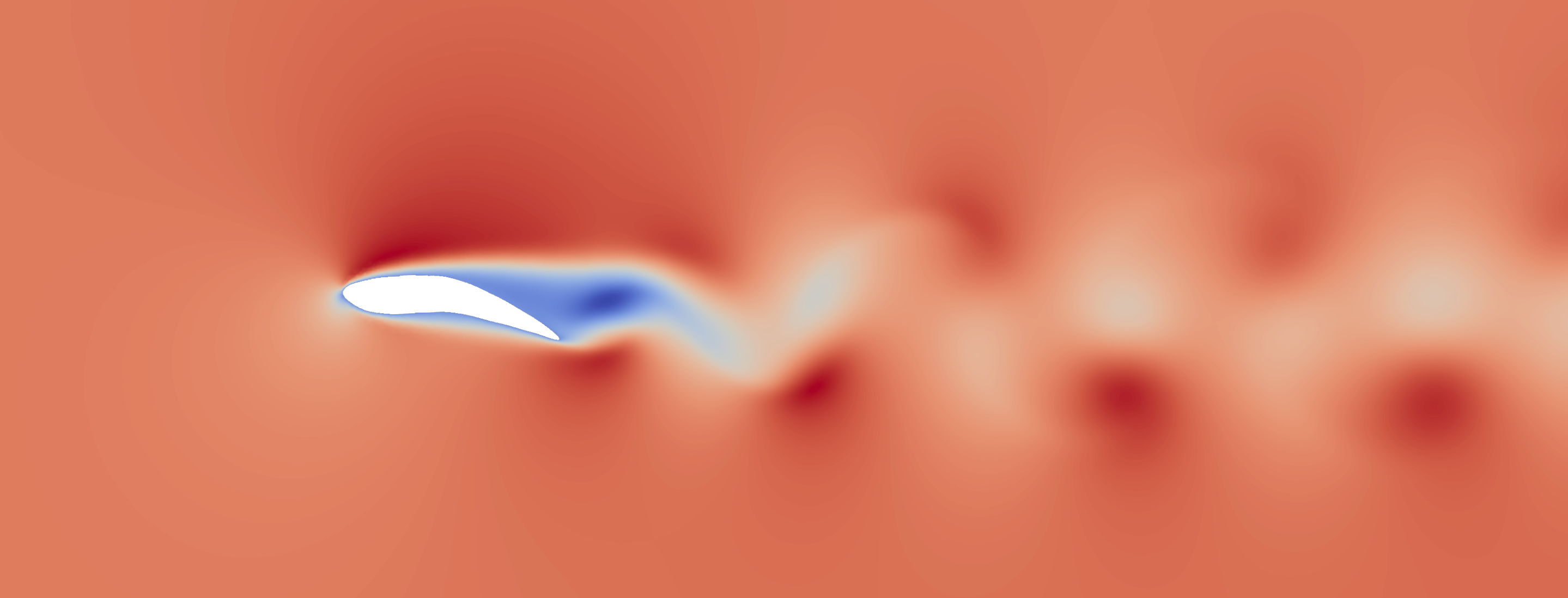}} 
	\caption{Computed $v_x$ velocity field at $Re \sim 600$ around shape \ref{fig:4_pts_4_pts} (the domain is cropped).}
	\label{fig:4_pts_4_pts_u}
\end{subfigure}
%%%%%%%
\caption{\textbf{Results of the baseline shape optimization process}. Best shapes obtained using 1, 3 and 4 free points are shown in subfigures \ref{fig:4_pts_1_pt}, \ref{fig:4_pts_3_pts} and \ref{fig:4_pts_4_pts} respectively. In subfigure \ref{fig:4_pts_1_pt}, the left, top and bottom points are fixed to their initial position (\ie that of the reference cylinder), while the rightmost one is free to move. In subfigure \ref{fig:4_pts_3_pts}, only the left point is fixed, while in subfigure \ref{fig:4_pts_4_pts}, all four points are free to move. The velocity field corresponding to shape \ref{fig:4_pts_4_pts} is shown in subfigure \ref{fig:4_pts_4_pts_u}.}
\label{fig:wing_4pts}
\end{figure}

\begin{figure}
\centering
%%%%%%%
\begin{subfigure}[t]{.25\textwidth}
	\centering
	\fbox{\includegraphics[width=\linewidth]{4_pts_4_pts.png}} 
	\caption{Best shape with 4 free points}
\end{subfigure} \qquad
\begin{subfigure}[t]{.25\textwidth}
	\centering
	\fbox{\includegraphics[width=\linewidth]{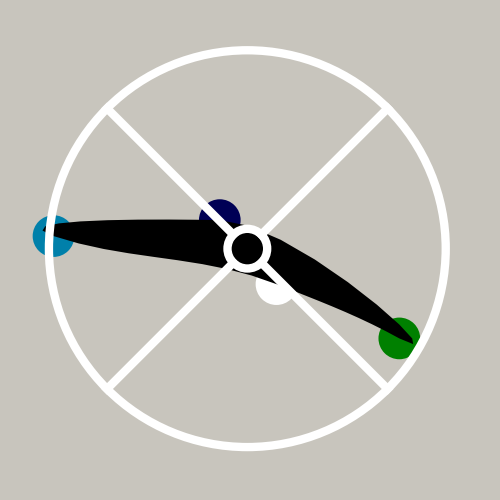}} 
	\caption{Close-to-best shape with 4 free points}
\end{subfigure} \qquad
\begin{subfigure}[t]{.25\textwidth}
	\centering
	\fbox{\includegraphics[width=\linewidth]{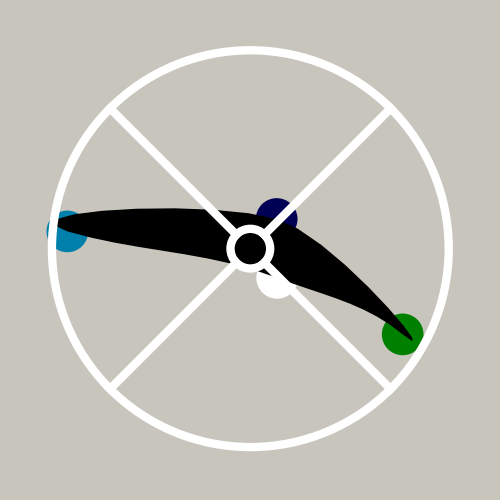}} 
	\caption{Close-to-best shape with 4 free points}
\end{subfigure}
%%%%%%%
\caption{\textbf{Some of the best performers using 4 free points.} A rounded leading edge is not an essential feature of the best performers. This arises from the low $Re$ value used in this experiment. By contrast, the trailing edge is similar in all the best-performing shapes.}
\label{fig:4_pts_leading_edge}
\end{figure}

\begin{figure}
\centering
%%%%%%%
\begin{subfigure}[t]{.45\textwidth}
	\centering
	\includegraphics[width=\linewidth]{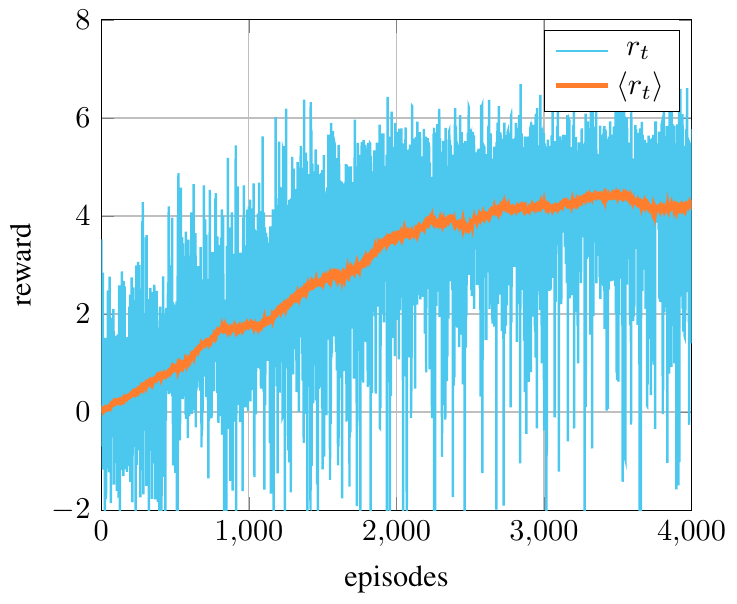}
	\caption{Instantaneous and moving-average reward history using 4 free points}
	\label{fig:reward_4_pts_4_pts}
\end{subfigure} \qquad
\begin{subfigure}[t]{.45\textwidth}
	\centering
	\includegraphics[width=.968\linewidth]{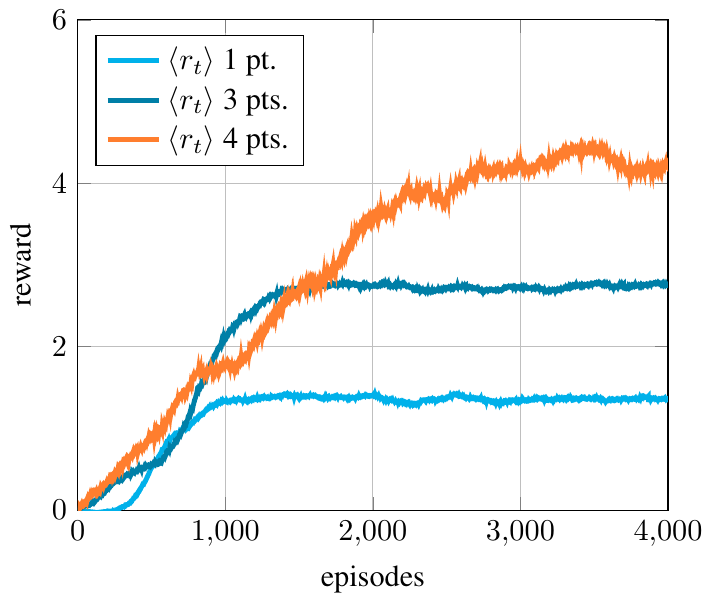}
	\caption{Moving-average reward history using 1, 3 and 4 free points}
	\label{fig:reward_averaged}
\end{subfigure}
%%%%%%%
\caption{\textbf{Typical reward evolution (instantaneous and averaged)} during the shape optimization process. Subfigure \ref{fig:reward_4_pts_4_pts} corresponds to the learning of case \ref{fig:4_pts_4_pts}. Moving-averaged learning curves for the three different cases of figure \ref{fig:wing_4pts} are compared in subfigure \ref{fig:reward_averaged}.}
\label{fig:wing_4pts_reward}
\end{figure}

The results obtained in this configuration with 1, 3 and
4 free points out of 4 points describing the shape in total
are shown in figure \ref{fig:wing_4pts}. As stated in section \ref{section:DRL},
one B\'ezier point corresponds to 3 degrees of freedom (d.o.f.) for the network
to optimize (the position of the point ($x,y$) and the local curvature $e$).
In the case of a single free
point (figure \ref{fig:4_pts_1_pt}), the agent understands the necessity of creating 
a high pressure area below 
the shape to generate lift, and creates an airfoil-like shape with a high angle of attack. 
The presence of a trailing edge is also observed on all the best-performing shapes. 
When using three free points (figure \ref{fig:4_pts_3_pts}, the same behavior 
is observed with a reduced apparent 
diameter, that is mostly controlled by the top and bottom points. The angle of attack 
is reduced compared to that of the single point case, while the average maximal 
reward is increased (see figure \ref{fig:reward_averaged}). When the four points are 
allowed to move (figure \ref{fig:4_pts_4_pts}), the airfoil extends to the whole 
available domain to maximize lift, but 
the trailing edge remains similar (in both 3 and 4 free points cases, the angle between 
the center of the shape and the trailing edge is close to \ang{23}). Although a rounded 
leading edge appears on the shape of figure \ref{fig:4_pts_4_pts}, it did not appear 
systematically in the other best-performing shapes, as shown in figure 
\ref{fig:4_pts_leading_edge}. This \tb{is most probably} due to the relatively low Reynolds number
used in the present study. Finally, it should be noted that with four available points, 
the agent manages to obtain even better rewards, as shown in figure 
\ref{fig:wing_4pts_reward}. In all cases, learning happens almost immediately, and 
continues almost linearly before reaching a plateau, after which the agent keeps 
exploring the environment, but no learning is specifically visible. The shapes shown in 
figures \ref{fig:4_pts_1_pt}, \ref{fig:4_pts_3_pts} and \ref{fig:4_pts_4_pts} are the best 
ones drawn from the whole exploration. The horizontal component of the velocity 
field around the last shape is also shown in figure \ref{fig:4_pts_4_pts_u}.

%%%%%%%%%%%%%%%%%%%%
%%%%%%%%%%%%%%%%%%%%

%\input{fig/wing_4pts}

%%%%%%%%%%%%%%%%%%%%
%%%%%%%%%%%%%%%%%%%%

%\input{fig/4_pts_leading_edge}

%%%%%%%%%%%%%%%%%%%%
%%%%%%%%%%%%%%%%%%%%

%\input{fig/wing_4pts_reward}

%%%%%%%%%%%%%%%%%%%%%%%%%%%%%%%%%%%%%%%%%%%%%%%%%%%%%%%%%%
%%%%%%%%%%%%%%%%%%%%%%%%%%%%%%%%%%%%%%%%%%%%%%%%%%%%%%%%%%
\subsection{Reward shaping}

%%%%%%%%%%%%%%%%%%%%%%%%%%%%%%%%%%%%%%%%%%%%%%%%%%%%%%%%%%%
\subsubsection{Shaping for faster convergence}

\begin{figure}
\centering
%%%%%%%
\begin{subfigure}[t]{.45\textwidth}
	\centering
	\includegraphics[width=\linewidth]{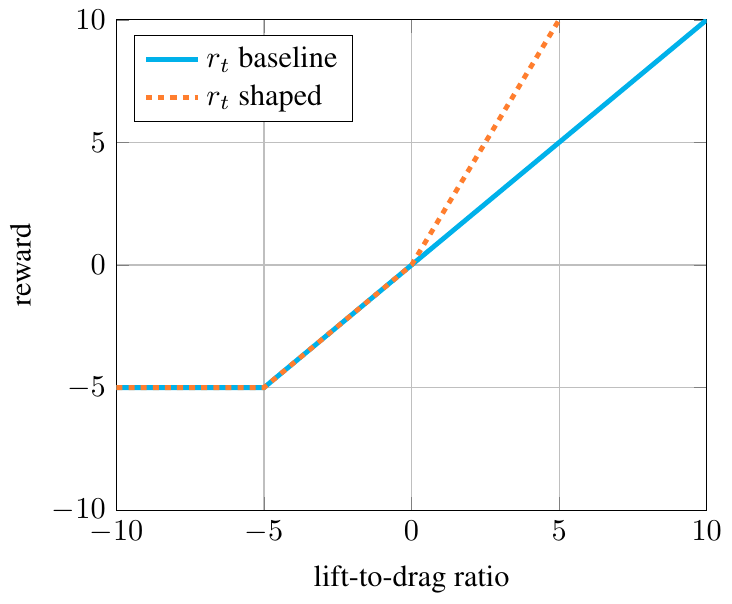}
	\caption{Baseline and shaped reward functions as a function of lift-to-drag ratio}
	\label{fig:reward_functions}
\end{subfigure} \qquad
\begin{subfigure}[t]{.45\textwidth}
	\centering
	\includegraphics[width=.975\linewidth]{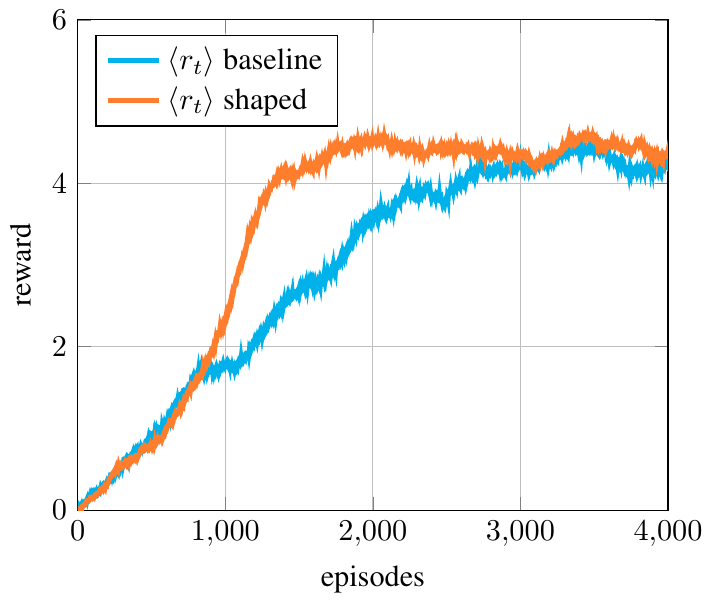}
	\caption{Average reward observed during agent learning using baseline and shaped reward functions (normalized)}
	\label{fig:reward_4_pts_4_pts_shaped}
\end{subfigure}
%%%%%%%
\caption{\textbf{Baseline and shaped reward functions and their resulting observed rewards} on the 4 free points case. Using a shaped reward increases the overall learning speed.}
\label{fig:reward_shaping}
\end{figure}

It can be observed on figure \ref{fig:reward_averaged} that the learning process requires a 
considerable amount of explored shapes to converge toward its final performance level. As could 
be expected, this number of shapes increases with the number of degrees of freedom involved in 
the shape generation. In this section, we show that basic reward shaping is enough, in 
our case, to cut that number by a considerable amount. To do so, the reward is computed 
following equations (\ref{eq:reward_wing}) and (\ref{eq:reward_penalize_wing}), after which 
it is multiplied by a constant if it is positive, as shown in figure \ref{fig:reward_functions},
following:

\begin{equation}
\label{eq:shaped_reward_wing}
	r_t \leftarrow 2 \, r_t \cdot \mathds{1} \big( r_t > 0 \big).
\end{equation} 

The impact of this modification on the learning is clearly visible in figure 
\ref{fig:reward_4_pts_4_pts_shaped}: when using the shaped reward, the agent reaches 
the learning plateau after approximatively 1500 shapes, versus 3000 when using the 
baseline reward. The average plateau reward is also slightly higher with the shaped reward.

%%%%%%%%%%%%%%%%%%%%
%%%%%%%%%%%%%%%%%%%%

%\input{fig/reward_shaping}

\newpage

%%%%%%%%%%%%%%%%%%%%%%%%%%%%%%%%%%%%%%%%%%%%%%%%%%%%%%%%%%%
\subsubsection{Shaping to add constraints}

\begin{figure}
\centering
%%%%%%%
\begin{subfigure}[t]{.3\textwidth}
	\centering
	\fbox{\includegraphics[width=\linewidth]{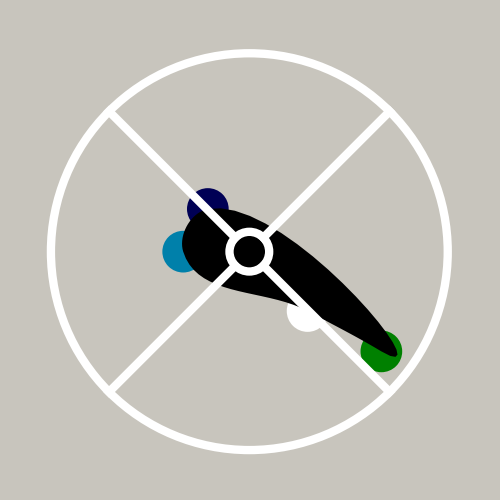}} 
	\caption{Best shape with 4 points, 3 free points, with area penalization ($\alpha = 3.176$)}
	\label{fig:4_pts_3_pts_area}
\end{subfigure} \qquad \qquad
\begin{subfigure}[t]{.3\textwidth}
	\centering
	\fbox{\includegraphics[width=\linewidth]{4_pts_3_pts.png}} 
	\caption{Best shape with 4 points, 3 free points, no area penalization ($\alpha = 2.733$)}
	\label{fig:4_pts_3_pts_no_area}
\end{subfigure}
%%%%%%%
\caption{\textbf{Optimal shapes obtained with and without area penalization} using 4 points with 3 free points. While the target area is that of the unit cylinder ($\alpha_{\text{cyl}} = \pi$), the area of the optimal shape with area penalization is equal to 3.176, versus 2.733 for that of the non-penalized shape. While respecting this constraint, the lift-to-drag ratio of the penalized shape is approximately 30\% lower than that of the non-penalized one.}
\label{fig:3pts_area}
\end{figure}

\begin{figure}
\centering
%%%%%%%
\begin{subfigure}[t]{.45\textwidth}
	\centering
	\includegraphics[width=.96\linewidth]{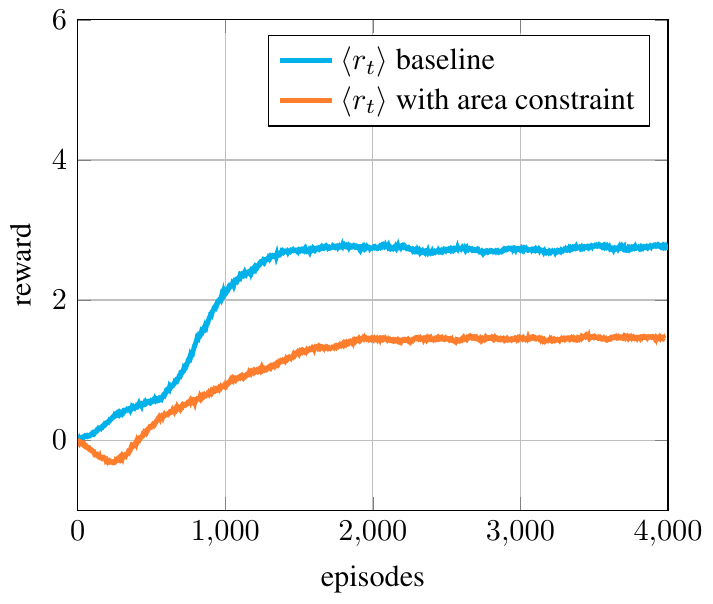}
	\caption{Moving-average reward history with and without area constraint}
	\label{fig:reward_area}
\end{subfigure} \qquad
\begin{subfigure}[t]{.45\textwidth}
	\centering
	\includegraphics[width=\linewidth]{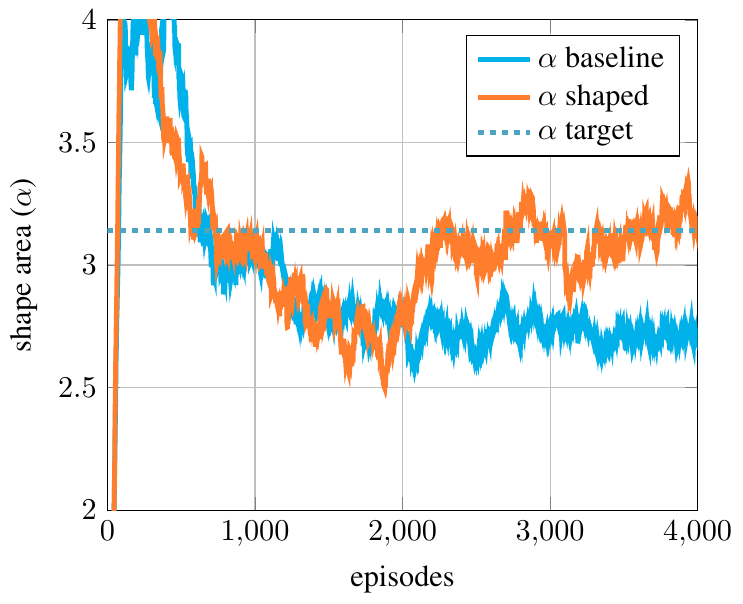}
	\caption{Moving-average area history with and without area constraint}
	\label{fig:area_area}
\end{subfigure}
%%%%%%%
\caption{\textbf{Moving-average reward and area history of the explored shapes using baseline and shaped reward (\ref{eq:shaped_reward_area}) compared to target area} on a single training. Once the learning plateau is reached (approx. 2000 episodes), the agent modifies its behavior in order to fulfill the additional area constraint, although this is barely visible on the reward curve. For the case with area constraint, the agent has to make a trade-off between the lift-to-drag ratio and the area penalization in the reward, explaining the lower $\left< r_t \right>$ values.}
\label{fig:reward_shaping_area}
\end{figure}

Constraints can be weakly enforced (in a non-mathematical sense) by adding penalties to 
the reward function that eventually prohibits undesired behaviors from the network. By regularly 
hitting a reward barrier in the action space, the agent will learn to avoid the associated 
behavior. Here, the goal is to prescribe the area of the optimal shape to remain close to that of 
the initial cylinder. To that end, one can simply add an \textit{ad hoc} penalization term 
to the reward function:

\begin{equation}
\label{eq:shaped_reward_area}
	r_t \leftarrow r_t - \frac{\left| \alpha - \alpha_{\text{cyl}} \right|}{\alpha_{\text{cyl}}},
\end{equation} 

where $\alpha$ is the area of the current shape, and $\alpha_{\text{cyl}}$ is the area 
of the reference cylinder. In figure \ref{fig:3pts_area}, we compare the optimal shapes 
obtained using 4 points with 3 moving points, both with and without the area penalization 
(\ref{eq:shaped_reward_area}). The area of the optimal shape with penalization is very close 
to that of the reference cylinder, which is not the case of the baseline one. As shown in 
figure \ref{fig:reward_shaping_area}, during the first 2000 episodes, both the constrained 
and unconstrained agents produce shapes of similar areas. Once the learning plateau is reached, 
the constrained agent starts generating shapes that minimize the 
penalization term in (\ref{eq:shaped_reward_area}). Although this effect is barely visible on the reward 
curves, the behavior is particularly clear when looking at the area history. 
The lower $\left< r_t \right>$ values for the constrained agent are a direct consequence of the trade-off 
between the lift-to-drag ratio and the area penalization.
This additional constraint induces a 
reduction of approximately 30\% in the lift-to-drag ratio compared to the optimal shape 
without area penalization.

%%%%%%%%%%%%%%%%%%%%
%%%%%%%%%%%%%%%%%%%%

%\input{fig/3pts_area}

%%%%%%%%%%%%%%%%%%%%
%%%%%%%%%%%%%%%%%%%%%

%\input{fig/reward_shaping_area}

\newpage

%%%%%%%%%%%%%%%%%%%%%%%%%%%%%%%%%%%%%%%%%%%%%%%%%%%%%%%%%%
%%%%%%%%%%%%%%%%%%%%%%%%%%%%%%%%%%%%%%%%%%%%%%%%%%%%%%%%%%
%%%%%%%%%%%%%%%%%%%%%%%%%%%%%%%%%%%%%%%%%%%%%%%%%%%%%%%%%%
\section{Conclusions}

In this work, we present the first application of deep reinforcement learning 
to direct shape optimization. After an introduction to the basic concepts of DRL 
and a description of the CFD setup, details were given on the shape generation using B\'ezier curves 
and the implementation of the DRL environment. Then, we showed that, provided 
an adequate reward function based on the lift-to-drag ratio, our agent generated wing-like 
optimal shapes without any priori knowledge of aerodynamic concepts. 
Furthermore, we explored reward shaping, both to speed up learning and to introduce additional 
constraint on the considered optimization problem.
This paper also introduced a "degenerate DRL" approach that allows to use DRL algorithms as 
general purpose optimizers.
Many refinements of the method remain to be explored, regarding the observations provided to 
the agent, the management of the shape deformation, or the data efficiency of the method, among others.

This contribution paves way to a new category of shape optimization processes. Although our application 
is focused on aerodynamics, the method is agnostic to the application use case, which makes our approach easily extendable 
to other domains of computational mechanics. In addition, using DRL for performing shape optimization may offer several promising
perspectives. First, the DRL methodology can be expected to handle non-linear, high dimensionality
optimization problems well, as this has already been proven in a number of control applications. Second, DRL
is known to scale well to large amounts of data, which is well adapted to cases where parallelization
of simulations is difficult due to algorithmic or hardware challenges, but many
simulations can be run in parallel. Third, we can expect that
transfer learning may enable DRL to solve similar new problems solely based on the knowledge obtained from its previous training.
Further work
should be performed to investigate each of these aspects.

%%%%%%%%%%%%%%%%%%%%%%%%%%%%%%%%%%%%%%%%%%%%%%%%%%%%%%%%%%
%%%%%%%%%%%%%%%%%%%%%%%%%%%%%%%%%%%%%%%%%%%%%%%%%%%%%%%%%%
%%%%%%%%%%%%%%%%%%%%%%%%%%%%%%%%%%%%%%%%%%%%%%%%%%%%%%%%%%
\section*{Acknowledgements}

This work was supported by the Carnot M.I.N.E.S MINDS project.

%%%%%%%%%%%%%%%%%%%%%%%%%%%%%%%%%%%%%%%%%%%%%%%%%%%%%%%%%%
%%%%%%%%%%%%%%%%%%%%%%%%%%%%%%%%%%%%%%%%%%%%%%%%%%%%%%%%%%
%%%%%%%%%%%%%%%%%%%%%%%%%%%%%%%%%%%%%%%%%%%%%%%%%%%%%%%%%%
\begin{appendices}

%%%%%%%%%%%%%%%%%%%%%%%%%%%%%%%%%%%%%%%%%%%%%%%%%%%%%%%%%%
%%%%%%%%%%%%%%%%%%%%%%%%%%%%%%%%%%%%%%%%%%%%%%%%%%%%%%%%%%
%%%%%%%%%%%%%%%%%%%%%%%%%%%%%%%%%%%%%%%%%%%%%%%%%%%%%%%%%%
\section{Open source code}

The code of this project is available on the following Github repository: 
\url{https://github.com/jviquerat/drl_shape_optimization}. It relies on 
FEniCS for the CFD resolution \cite{Fenics2015}, and on Tensorforce 
\cite{tensorforce} for the reinforcement learning library. The generation of 
shapes using B\'ezier curves description is ensured by a homemade code 
included in the repository.

%%%%%%%%%%%%%%%%%%%%%%%%%%%%%%%%%%%%%%%%%%%%%%%%%%%%%%%%%%
%%%%%%%%%%%%%%%%%%%%%%%%%%%%%%%%%%%%%%%%%%%%%%%%%%%%%%%%%%
%%%%%%%%%%%%%%%%%%%%%%%%%%%%%%%%%%%%%%%%%%%%%%%%%%%%%%%%%%
\section{DRL, policy gradient and PPO algorithm}
\label{appendix:ppo}

This section briefly introduces the basic concepts and definitions of DRL, policy gradient 
and proximal policy optimization methods. For a more detailed introduction and discussion,
the reader is referred to \cite{Francois-lavet2018} and references therein for more details.

Reinforcement learning is a class of machine learning methods focusing on
optimal decision making in a complex environment. At any discrete timestep
$t \in \mathbb{N}$, an agent observes the current world state $s_t$, decides 
for an action $a_t$ and receives a reward signal $r_t \in \mathbb{R}$. In
the literature, observation and state are sometimes distinguished, but for ease of notation,
these two concepts are commonly merged into the concept of a state $s_t$ (which we do here).
Still, the reader must keep in mind that states are often
a partial and/or noisy observation of the actual state of the environment. The final goal of
the agent is to update the discounted cumulative reward over a rollout of the agent's policy $\pi$,
\ie a trajectory of states, actions and rewards $\tau = (s_0, a_0, r_0 s_1, a_1, r_1, ...)$ which 
distribution follows policy $\pi$:

\begin{equation*}
	R(\tau)= \sum_{t=0}^T \gamma^t  r_{t},
\end{equation*} 

\noindent where $\gamma \in \left[0,1\right]$ is a discount factor to prioritize more immediate
over more distant rewards. Two popular types of 
reinforcement learning algorithms are Q-learning and policy gradient methods:

\begin{enumerate}
	\item Q-learning assumes a discrete, finite action space, and chooses actions 
	based on their estimated Q-value, which is the expected discounted cumulative 
	reward obtained when starting from state $s$ with action $a$, and then 
        following trajectory $\tau$ according to policy $\pi$:
	
	\begin{equation*}
		Q(s, a) = \underset{\tau \sim \pi}{\mathbb{E}} \left[ R(\tau) \vert s, a \right].
	\end{equation*} 
	
	In DRL, the $Q$ function is implemented as a deep neural network and optimized 
	to fit the recursively characterized optimal solution, given by the Bellman equation:
	
	\begin{equation*}
		Q^*(s, a) = R(s,a) + \gamma \max_{a'} Q^*(s', a').
	\end{equation*} 

	\item Policy gradient (PG) methods, on the other hand, can handle both discrete and 
        continuous action spaces. In contrast to Q-learning, PG methods directly optimize 
        the policy instead of an auxiliary value function. They assume a
        stochastic policy $\pi(a \vert s)$, 
	often parameterized by a deep neural network, whose gradient-based optimization 
	directly maximizes the expected discounted cumulative reward
    	$\underset{\tau \sim \pi}{\mathbb{E}} \left[ R \right]$, approximated
	by a finite batch of rollouts. Compared to Q-learning methods, PG methods exhibit
	better capabilities in handling high dimensional action spaces as well as smoother 
	convergence properties, although they are known to often converge to local minima.
\end{enumerate}

Introduced in 2000 by Sutton \etal \cite{Sutton2000}, vanilla PG relies on an estimate of the 
first-order gradient of the log-policy $\nabla_{\theta} \log \pi_{\theta}$ to update its network. 
This approach was latter followed by several major improvements, including the trust-region 
policy optimization (TRPO) \cite{Schulman2015} and the proximal policy optimization (PPO) 
\cite{Schulman2017}. In these methods, the network update exploits a 
\emph{surrogate advantage} functional:

\begin{equation*}
	\theta_{k+1} = \arg \max_{\theta} L(\theta_k, \theta),
\end{equation*}

\noindent with

\begin{equation*}
	L(\theta_k, \theta) = \underset{(s,a) \sim \pi_{\theta_k}}{\mathbb{E}} \big[ \Pi(s,a,\theta,\theta_k) A^{\pi_{\theta_k}} (s,a) \big],
\end{equation*}

\noindent and

\begin{equation*}
	\Pi(s, a, \theta, \theta_k) = \frac{\pi_\theta (a \vert s)}{\pi_{\theta_k} (a \vert s)}.
\end{equation*}
	
In the latter expressions, $A^{\pi_{\theta_k}} (s,a)$ is called the advantage function, and measures 
how much better it is to take action $a$ in state $s$ compared to the average result of all actions 
that could be taken in state $s$. Hence, $L(\theta_k, \theta)$ measures how much better (or worse) 
policy $\pi_{\theta}$ performs compared to the previous policy $\pi_{\theta_k}$. In order to avoid too
large policy updates that could collapse the policy performance, TRPO leverages second-order 
natural gradient optimization to update parameters within a trust-region of a fixed maximum 
Kullback-Leibler divergence between old and updated policy distribution. This relatively
complex approach was replaced in the PPO method by simply clipping the maximized
expression:

\begin{equation*}
	L(\theta_k, \theta) = \underset{(s,a) \sim \pi_{\theta_k}}{\mathbb{E}} \big[ \min \left( \Pi(s,a,\theta,\theta_k) A^{\pi_{\theta_k}} (s,a), g\left( \eps, A^{\pi_{\theta_k}} (s,a) \right) \right) \big],
\end{equation*}

\noindent where

\begin{equation*}
	g(\eps, A) = 
	\begin{cases}
		(1+\eps) A 	& A \geq 0,\\
		(1-\eps) A 		& A < 0,
	\end{cases}
\end{equation*}

\noindent where $\eps$ is a small, user-defined parameter. When $A^{\pi_{\theta_k}} (s,a)$ is positive, 
taking action $a$ in state $s$ is preferable to the average of all actions that could be taken in that state, 
and it is natural to update the policy to favor this action. Still, if the ratio $\Pi(s,a,\theta,\theta_k)$ is very large, 
stepping too far from the previous policy $\pi_{\theta_k}$ could damage performance. For that reason, 
$\Pi(s,a,\theta,\theta_k)$ is clipped to $1+\eps$ to avoid too large updates of the policy. If 
$A^{\pi_{\theta_k}} (s,a)$ is negative, taking action $a$ in state $s$ represents a poorer choice than 
the average of all actions that could be taken in that state, and it is natural to update the policy to 
decrease the probability of taking this action. In the same fashion, $\Pi(s,a,\theta,\theta_k)$ is clipped 
to $1-\eps$ if it happens to be lower than that value. 

In the latter expressions, $A^{\pi_{\theta_k}} (s,a)$ is estimated using a generalized advantage 
estimator (GAE), which represents a trade-off between Monte-Carlo and time difference estimators 
\cite{Schulman2015b}. Additionally, instead of performing a single, full-batch update, the network 
optimization is decomposed in multiple updates computed from subsampled mini-batches. Finally, 
an entropy regularization is added to the surrogate loss:

\begin{equation*}
	L(\theta_k, \theta) = \underset{(s,a) \sim \pi_{\theta_k}}{\mathbb{E}} \big[ \min \left( \Pi(s,a,\theta,\theta_k) A^{\pi_{\theta_k}} (s,a), g\left( \eps, A^{\pi_{\theta_k}} (s,a) \right) \right) \big] + c H \left( \pi_{\theta} (a \vert s) \right).
\end{equation*}

This additional term encourages the agent not to be over-confident, by keeping the
policy distribution close to uniform unless there is a strong signal not to.

%%%%%%%%%%%%%%%%%%%%%%%%%%%%%%%%%%%%%%%%%%%%%%%%%%%%%%%%%%
%%%%%%%%%%%%%%%%%%%%%%%%%%%%%%%%%%%%%%%%%%%%%%%%%%%%%%%%%%
%%%%%%%%%%%%%%%%%%%%%%%%%%%%%%%%%%%%%%%%%%%%%%%%%%%%%%%%%%
\section{Shape generation using B\'ezier curves}
\label{appendix:shapes}

\begin{figure}
\centering
%%%%%%%
\begin{subfigure}[t]{.45\textwidth}
	\centering
	\includegraphics[width=\linewidth]{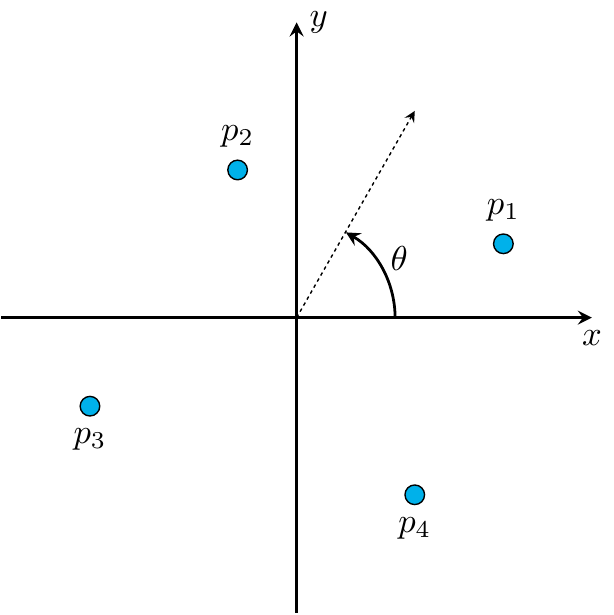}
	\caption{Sort the provided points by ascending trigonometric angle}
	\label{fig:shape_generation_1}
\end{subfigure} \qquad
\begin{subfigure}[t]{.45\textwidth}
	\centering
	\includegraphics[width=\linewidth]{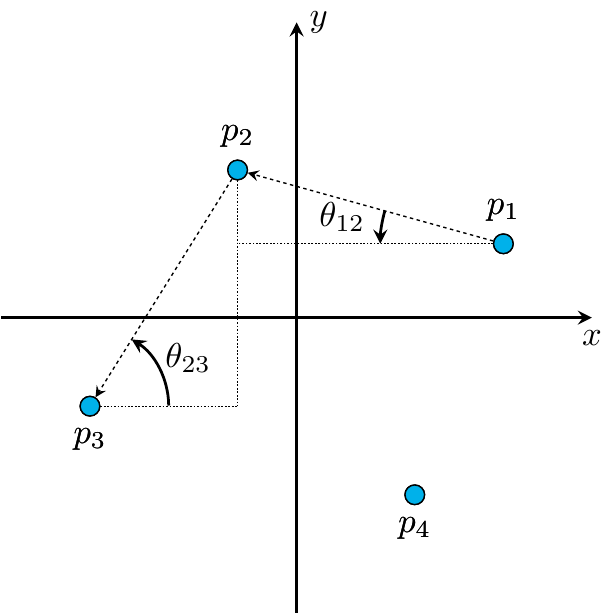}
	\caption{Compute angles between points, and compute an average angle around each point $\theta^*_i$}
	\label{fig:shape_generation_2}
\end{subfigure}

\medskip
\medskip

\begin{subfigure}[t]{.45\textwidth}
	\centering
	\includegraphics[width=\linewidth]{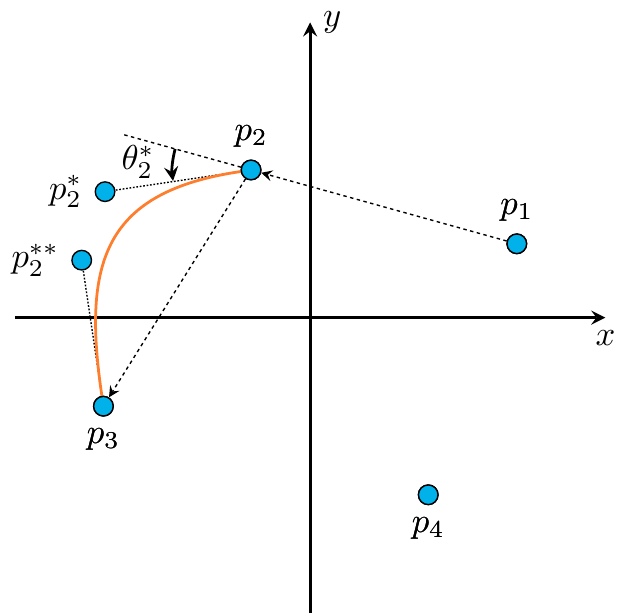}
	\caption{Compute control points coordinates from averaged angles and generate cubic B\'ezier curve}
	\label{fig:shape_generation_3}
\end{subfigure} \qquad
\begin{subfigure}[t]{.45\textwidth}
	\centering
	\includegraphics[width=\linewidth]{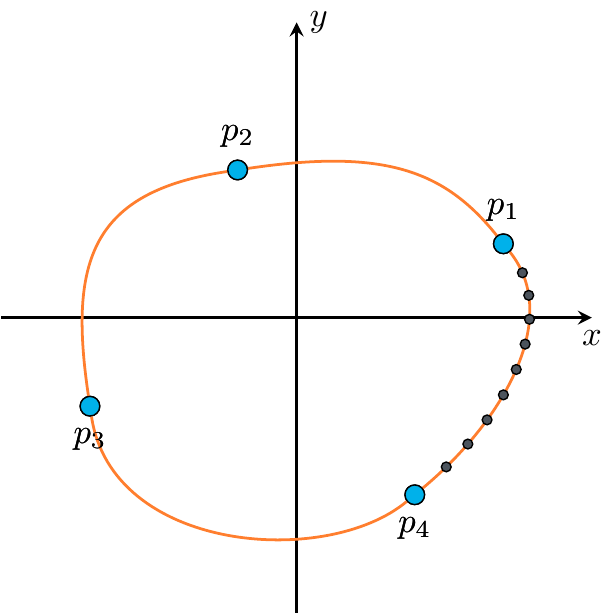}
	\caption{Sample all B\'ezier lines and export for mesh immersion}
	\label{fig:shape_generation_4}
\end{subfigure}
%%%%%%%
\caption{\textbf{Shape generation using cubic B\'ezier curves}. Each subfigure illustrates one of the consecutive steps used in the process. Detailed nomenclature is provided in the text.}
\label{fig:shape_generation}
\end{figure}

This section describes the process followed to generate shapes from a set 
of $n_s$ points provided by the agent. Once the points are collected, an ascending 
trigonometric angle sort is performed (see figure \ref{fig:shape_generation_1}), 
and the angles between consecutive points are computed. An average 
angle is then computed around each point 
(see figure \ref{fig:shape_generation_2}) using:

\begin{equation*}
	\theta^*_i = \alpha \theta_{i-1,i} + (1 - \alpha) \theta_{i,i+1},
\end{equation*}

\noindent with $\alpha \in \left[0,1\right]$. The averaging parameter $\alpha$ allows to 
alter the sharpness of the curve locally, maximum smoothness being obtained for 
$\alpha = 0.5$. Then, each pair of points is joined using a cubic B\'ezier curve, 
defined by four points: the first and last points, $p_i$ and $p_{i+1}$, are part of the curve, 
while the second and third ones, $p^*_i$ and $p^{**}_i$, are control points that define 
the tangent of the curve at $p_i$ and $p_{i+1}$. The tangents at $p_i$ and $p_{i+1}$ are 
respectively controlled by $\theta^*_i$ and $\theta^*_{i+1}$ (see figure \ref{fig:shape_generation_3}). 
A final sampling of the successive B\'ezier curves leads to a boundary description 
of the shape (figure \ref{fig:shape_generation_4}). Using this method, a wide variety of shapes can be attained.

%%%%%%%%%%%%%%%%%%%%
%%%%%%%%%%%%%%%%%%%%

%\input{fig/shape_generation}

\end{appendices}

\newpage

%%%%%%%%%%%%%%%%%%%%%%%%%%%%%%%%%%%%%%%%%%%%%%%%%%%%%%%%%%%
%%%%%%%%%%%%%%%%%%%%%%%%%%%%%%%%%%%%%%%%%%%%%%%%%%%%%%%%%%%
\bibliographystyle{unsrt}
\bibliography{nn_flow_2d}

\end{document}